\documentclass[a4paper]{spie}%
\usepackage{graphics}
\usepackage{amsmath}
\usepackage{graphicx}
\usepackage{amsfonts}
\usepackage{amssymb}%
\authorinfo{Further author information: Send correspondence to G.S. (gae.spedalieri@york.ac.uk)}
\begin{document}

\title{Quantum cryptography with an ideal local relay}
\author{Gaetana Spedalieri\supit{a}, Carlo Ottaviani\supit{a}, Samuel L.
Braunstein\supit{a}, Tobias Gehring\supit{b}, Christian S. Jacobsen\supit{b},
Ulrik L. Andersen\supit{b}, and Stefano Pirandola\supit{a}
\skiplinehalf \supit{a}Computer Science and York Centre for Quantum
Technologies,University of York, Deramore Lane, York YO10 5GH, United Kingdom
\skiplinehalf
\supit{b}Department of Physics, Technical University of Denmark, Fysikvej,
2800 Kongens Lyngby, Denmark}
\maketitle

\begin{abstract}
We consider two remote parties connected to a relay by two quantum
channels. To generate a secret key, they transmit coherent states
to the relay, where the states are subject to a
continuous-variable (CV) Bell detection. We study the ideal case
where Alice's channel is lossless, i.e., the relay is locally
situated in her lab and the Bell detection is performed with unit
efficiency. This configuration allows us to explore the optimal
performances achievable by CV measurement-device-independent
quantum key distribution. This corresponds to the limit of a
trusted local relay, where the detection loss can be re-scaled.
Our theoretical analysis is confirmed by an experimental
simulation where $10^{-4}$ secret bits per use can potentially be
distributed at $170$km assuming ideal reconciliation.

\end{abstract}

\section{Introduction}

Quantum key distribution (QKD)~\cite{Gisin,Scarani} is a central area in
quantum information science~\cite{Wilde,RMP}. A typical QKD protocol involves
two parties, Alice and Bob, who generate secret keys by exchanging quantum
systems over an insecure communication channel. Another scenario involves a
swapping-like protocol~\cite{SidePRL} where secret correlations are
established by the measurement of a third untrusted party (relay). This idea
of `measurement-device independence'
(MDI)\cite{SidePRL,Lo,Others,Oth1,Oth2,Oth3,Oth4,Oth4b,Oth5,EXP1,EXP2}\ has
been extended to continuous-variable (CV) systems\cite{CVMDI1,CVMDI2}, with
the possibility of much higher key rates.

In this paper, we study consider a limit configuration for CV-MDI-QKD, where
the relay is in Alice's lab and performs an ideal Bell detection. This is an
extrapolation which allows us to investigate the maximal rate/distance
performances achievable by CV-MDI-QKD. Experimentally, this is equivalent to
consider a local relay where the loss associated with the various technical
imperfections (such as the detector inefficiencies) can be re-scaled and
therefore neglected. As a matter of fact, this limit case corresponds to the
case where the loss of the relay is trusted.

Our theoretical analysis, confirmed by an experimental simulation, shows that
$10^{-2}$ secret bits per relay use can be distributed at $10$dB loss in Bob's
channel, equivalent to $50$km of standard optical fibre (at the loss rate of
$0.2$dB/km). Assuming ideal reconciliation, a potential rate of about
$10^{-4}$ secret bits per relay use can be distributed over a very lossy link,
i.e., $34$dB loss corresponding to $170$km of fibre.

\section{Protocol}

\begin{figure}[ptbh]
\vspace{-2.0cm}
\par
\begin{center}
\includegraphics[width=0.62\textwidth] {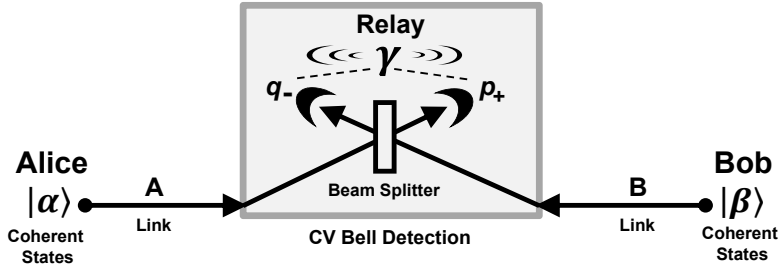}
\end{center}
\par
\vspace{-2.2cm}\caption{Basic protocol. See text for explanations.}%
\label{PMscheme}%
\end{figure}

The scheme is depicted in Fig.~\ref{PMscheme}. At one side, Alice prepares a
mode $A$ in a coherent state $\left\vert \alpha\right\rangle $ with
Gaussian-modulated amplitude $\alpha$; at the other side, Bob prepares mode
$B$ in another coherent state $\left\vert \beta\right\rangle $ with
Gaussian-modulated amplitude $\beta$ (Gaussian distributions have zero mean
and large variance).\ Modes $A$ and $B$ are sent to the relay, which performs
a CV Bell detection~\cite{BellFORMULA}, by mixing the modes in a balanced beam
splitter whose output ports are conjugately homodyned with outputs $q_{-}$ and
$p_{+}$. The complex variable $\gamma:=(q_{-}+ip_{+})/\sqrt{2}$ is then
communicated to Alice and Bob via a classical public channel. Since
$\gamma\simeq\alpha-\beta^{\ast}$, each party may infer the variable of the
other party by postprocessing. 

The most general eavesdropping strategy is a joint attack involving both the
relay and the two links\cite{CVMDI1,CVMDI2}. Here we consider the simple case
where Eve attacks Bob's link $B$ only, by means of a Gaussian
attack~\cite{canATTACKS} which introduces loss and thermal noise. The
travelling mode $B$ is mixed with an ancillary mode $E$ by a beam splitter
with transmissivity $\tau$. The ancilla introduces thermal noise with variance
$\omega$ and belongs to a reservoir of ancillas under Eve's control. This kind
of entangling-cloner attack~\cite{RMP} is repeated for each use of the relay
and the output ancillas are finally detected by Eve by means of an optimized
collective quantum measurement.

\section{Secret-key rate}

\begin{figure}[ptbh] \vspace{-3.0cm}
\par
\begin{center}
\includegraphics[width=0.62\textwidth] {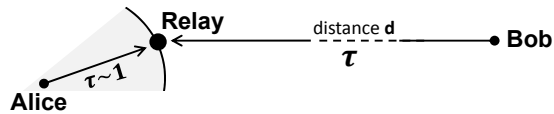}
\end{center}
\par
\vspace{-3.0cm}\caption{Key-distribution via an ideal local relay.
This is locally in Alice's lab and it is assumed to work perfectly
(unit quantum efficiency). By contrast, Bob's link has
transmissivity $\tau<1$ corresponding to some distance $d$ in
standard optical fibre. Bob's channel can also be
affected by thermal/excess noise.}%
\label{radius}%
\end{figure}

By specializing the formulas of CV-MDI-QKD\cite{CVMDI1}, we can derive the
secret-key rate for the scenario depicted in Fig.~\ref{radius}. Assuming ideal
reconciliation efficiency and large modulation, the key rate is given by%
\begin{equation}
R=h\left(  \tfrac{\chi}{1+\tau}-1\right)  -h\left[  \tfrac{\tau\chi
-(1+\tau)^{2}}{1-\tau^{2}}\right]  +\log_{2}\left[  \tfrac{2(1+\tau)}%
{e(1-\tau)\chi}\right]  ,\label{RateASYMM}%
\end{equation}
where%
\begin{equation}
h(x):=\frac{x+1}{2}\log_{2}\frac{x+1}{2}-\frac{x-1}{2}\log_{2}\frac{x-1}{2},
\end{equation}
and $\chi$ is the equivalent noise, decomposable as $\chi=\chi_{\text{loss}%
}+\varepsilon$, where $\chi_{\text{loss}}=2(1+\tau)/\tau$ is the noise due to
loss, while $\varepsilon$ is the `excess noise'. The maximum theoretical
performance of the protocol, with respect to the loss present in Bob's link,
is reached for $\varepsilon=0$. In this case, we have
\begin{equation}
R_{\text{loss}}=h[(2-\tau)/\tau]+\log_{2}[\tau/(1-\tau)e],
\end{equation}
which goes to zero only for $\tau\rightarrow0$, corresponding to
Bob arbitrarily far from the relay. It is easy to convert
transmissivity $\tau$ to distance $d$ in optical fibre, by
considering the standard loss rate of
$0.2$dB/km.

\begin{figure}[ptbh]
\vspace{-0.5cm}
\par
\begin{center}
\includegraphics[width=0.55\textwidth] {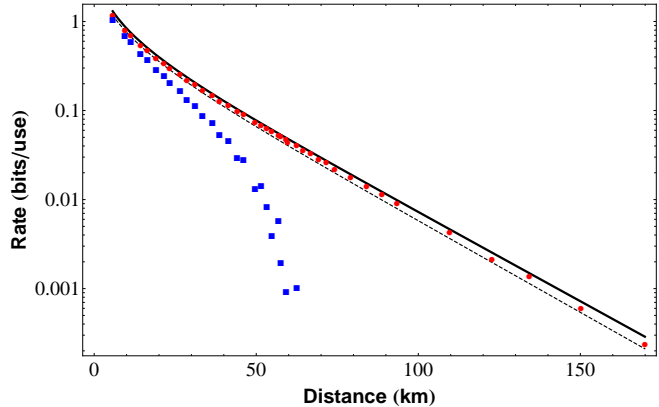}
\end{center}
\par
\vspace{-0.3cm}\caption{Secret-key rate $R$ versus Bob's distance
$d$ from the relay. Experimental points refer to ideal
reconciliation ($\xi=1$, red circles) and realistic reconciliation
($\xi\simeq0.97$, blue squares). For comparison, we also plot the
theoretical rates for a pure-loss attack (solid line) and a
Gaussian attack with excess noise $\varepsilon=0.02$ (dashed
line).}%
\label{picEXP}%
\end{figure}

In the scenario of Fig.~\ref{radius}, Bob can be very far from the relay also
in the presence of non-zero excess noise $\varepsilon\neq0$, with potential
distances beyond $100~$km of simulated fibre. This can be seen from the
numerical results shown in Fig.~\ref{picEXP}, where the solid line represents
the case of a pure-loss attack ($\varepsilon=0$), while the dashed curve
corresponds to an attack with non-zero excess noise, in particular
$\varepsilon=0.02$. We can see the robustness of the key rate with respect to
the excess noise.

This theoretical analysis is also confirmed by a
proof-of-principle experiment where we have realized the local
ideal relay by suitably re-scaling the loss in Alice's link in the
post-processing of the data. We have reproduced the extreme
asymmetric configuration of Fig.~\ref{radius}, with variable Bob's
transmissivity $\tau$, down to $4\times10^{-4}$ corresponding to
about $170$km in standard optical fibre. For every experimental
point, we have evaluated the key rate $R$ assuming ideal
reconcilation efficiency $\xi=1$. Experimental results are plotted
in Fig.~\ref{picEXP}\ and compared with the theoretical
predictions, with excellent agreement. The extrapolated
experimental rate approaches the theoretical limit of the
pure-loss attack. Due to imperfections, we have an excess noise
$\varepsilon\lesssim0.02$. Note that we can potentially reach
$R\simeq10^{-4}$ secret bits per relay use over a link with $34$dB
loss, equivalent to $170$km of optical fibre. 

Note that the current reconciliation procedures for CV protocols do not have
unit efficiency (indeed this is one of the main factors limiting the distance
of CV-QKD). By taking this limitation into account\cite{Jouguet} ($\xi
\simeq0.97$), we can still reach remarkably high rates over distances well
beyond the typical connection lengths of a network. As we can see from
Fig.~\ref{picEXP}, one can potentially achieve $R\simeq10^{-2}$ secret bits
per relay use over a link with $10$dB loss, equivalent to $50$km of optical fibre.

\section{Conclusion}

In this work, we have explored the maximal performances in terms of rates and
distances achievable by CV-MDI-QKD with coherent states. We have considered
the extreme configuration where the relay is in Alice's lab and the Bell
detection is ideally performed. An important feature of this protocol is the
simplicity of the relay, which does not possess any quantum source but just
performs a standard optical measurement, with all the heavy procedures of data
post-processing left to the end-users, fulfilling the idea behind the
end-to-end principle\cite{endtoend}. CV Bell detection involves highly
efficient photodetectors plus linear optics, whereas the discrete-variable
version of this measurement needs nonlinear elements to operate
deterministically. This feature combined with the use of coherent states makes
the scheme very attractive, guaranteeing both cheap implementation and
extremely high rates.

Our study also shows how improvements in the classical reconciliation
techniques (from $\xi\simeq0.97$ to $\xi=1$) have a dramatic impact on the
performances of the protocol, which means that the development of more
efficient classical codes for error correction and privacy amplification is a
central task in CV-QKD. Finally, future investigations could involve the
explicit security analysis of mixed technology environments where some of the
connections are established at low frequencies (infrared or microwave) where
thermal effects become important~\cite{Weed2010,Weed2012,Weed2013}.

\section*{Acknowledgments}

This work was supported by EPSRC (Grants EP/J00796X/1 and EP/L011298/1) and
the Leverhulme Trust.


\begin{thebibliography}{99}                                                                                               %


\bibitem {Gisin}Gisin, N., Ribordy, G., Tittel, W. \& Zbinden, H. Quantum
cryptography. \textit{Rev. Mod. Phys.} \textbf{74}, 145 (2002).

\bibitem {Scarani}Scarani, V., Bechmann-Pasquinucci, H., Cerf, N. J., Dusek,
M., Lutkenhaus, N. \& Peev, M. The security of practical quantum key
distribution. \textit{Rev. Mod. Phys.} \textbf{81}, 1301 (2009).

\bibitem {Wilde}Wilde, M. M. \textit{Quantum Information Theory} (Cambridge
University Press, Cambridge, 2013).

\bibitem {RMP}Weedbrook, C., Pirandola, S., Garcia-Patron, R., Cerf, N. J.,
Ralph, T. C., Shapiro, J. H. \& Lloyd, S. Gaussian quantum information.
\textit{Rev. Mod. Phys.} \textbf{84}, 621 (2012).

\bibitem {SidePRL}Braunstein, S. L. \& Pirandola, S. Side-channel-free quantum
key distribution. \textit{Phys. Rev. Lett.} \textbf{108}, 130502 (2012).

\bibitem {Lo}Lo, H.-K., Curty, M. \& Qi, B. Measurement-device-independent
quantum key distribution. \textit{Phys. Rev. Lett.} \textbf{108}, 130503 (2012).

\bibitem {Oth1}Ma, X., Fred Fung, C.-H. \& Razavi, M. Statistical fluctuation
analysis for measurement-device-independent quantum key distribution.
\textit{Phys. Rev. A} \textbf{86}, 052305 (2012).

\bibitem {Oth3}Ma, X. \& Razavi, M. Alternative schemes for
measurement-device-independent quantum key distribution. \textit{Phys. Rev. A}
\textbf{86}, 062319 (2012).

\bibitem {Others}Wang, X.B. Three-intensity decoy state method for device
independent quantum key distribution with basis dependent errors.
\textit{Phys. Rev. A} \textbf{87}, 012320 (2013).

\bibitem {Oth2}Branciard, C., Rosset, D., Liang, Y.-C. \& Gisin, N.
Measurement-device-independent entanglement witnesses for all entangled
quantum states. \textit{Phys. Rev. Lett.} \textbf{110}, 060405 (2013).

\bibitem {Oth4}Tomamichel, M., Fehr, S., Kaniewski, J. \& Wehner, S. A
monogamy-of-entanglement game with applications to device-independent quantum
cryptography. \textit{New J. Phys.} \textbf{15}, 103002 (2013).

\bibitem {Oth4b}Ci Wen Lim, C., Portmann, C., Tomamichel, M., Renner, R. \&
Gisin, N. Device-independent quantum key distribution with local Bell test.
\textit{Phys. Rev. X} \textbf{3}, 031006 (2013).

\bibitem {Oth5}Abruzzo, S., Kampermann, H., \& Bru\ss \ D.
Measurement-device-independent quantum key distribution with quantum memories.
\textit{Phys. Rev. A} \textbf{89}, 012301 (2014).

\bibitem {EXP1}Rubenok, A., Slater, J. A., Chan, P., Lucio-Martinez, I. \&
Tittel, W. Real-World two-photon interference and proof-of-principle quantum
key distribution immune to detector attacks. \textit{Phys. Rev. Lett.}
\textbf{111}, 130501 (2013).

\bibitem {EXP2}Ferreira da Silva, T., Vitoreti, D., Xavier, G. B., do Amaral,
G. C., Tempor\~{a}o, G. P. \& von der Weid, J. P. Proof-of-principle
demonstration of measurement-device-independent quantum key distribution using
polarization qubits. \textit{Phys. Rev. A} \textbf{88}, 052303 (2013).

\bibitem {CVMDI1}Pirandola, S., et al. High-rate
measurement-device-independent quantum cryptography. Nature Photon.
\textbf{9}, 397--402 (2015).

\bibitem {CVMDI2}Ottaviani, C., Spedalieri, G., Braunstein, S. L. \&
Pirandola, S. Continuous-variable quantum cryptography with an untrusted
relay: Detailed security analysis of the symmetric configuration. Phys. Rev. A
\textbf{91}, 022320 (2015).

\bibitem {BellFORMULA}Spedalieri, G., Ottaviani, C. \& Pirandola, S.
Covariance matrices under Bell-like detections. \textit{Open Syst. Inf. Dyn.}
\textbf{20}, 1350011 (2013).

\bibitem {canATTACKS}Pirandola, S., Braunstein, S. L. \& Lloyd, S.
Characterization of collective Gaussian attacks and security of coherent-state
quantum cryptography. \textit{Phys. Rev. Lett.} \textbf{101}, 200504 (2008).

\bibitem {Jouguet}Jouguet, P., Kunz-Jacques, S., \& Leverrier, A.
Long-distance continuous-variable quantum key distribution with a Gaussian
modulation. \textit{Phys. Rev. A} \textbf{84}, 062317 (2011).

\bibitem {endtoend}Saltzer, J. H., Reed, D. P. \& Clark, D. D. End-to-end
arguments in system design. \textit{Proceedings of the Second International
Conference on Distributed Computing Systems} (Paris, France, April 8-10, 1981).

\bibitem {Weed2010}Weedbrook, C., Pirandola, S., Lloyd, S. \& Ralph, T.~C.
Quantum cryptography approaching the classical limit. \textit{Phys. Rev.
Lett.} \textbf{105}, 110501 (2010).

\bibitem {Weed2012}Weedbrook, C., Pirandola, S. \& Ralph, T.~C.
Continuous-variable quantum key distribution using thermal states.
\textit{Phys. Rev. A} \textbf{86}, 022318 (2012).

\bibitem {Weed2013}Weedbrook, C., Ottaviani, C., \& Pirandola, S. Two-way
quantum cryptography at different wavelengths. \textit{Phys. Rev. A}
\textbf{89}, 012309 (2014).
\end{thebibliography}
\end{document}